\documentclass[doublecol]{epl2}

\usepackage{epstopdf}
\usepackage{amsmath}
\usepackage{amssymb}

\title{ Faraday Rotation and One-Way Propagation of Plasmon Waves on a Nanoparticle Chain}
\shorttitle{Faraday Rotation and One-Way Waves} 

\author{N. A. Pike\inst{1} \and D. Stroud\inst{1}}
\shortauthor{Pike and Stroud}

\institute{                    
  \inst{1} Department of Physics, The Ohio State University, Columbus, OH, USA 43210}


\abstract{We calculate the dispersion relations of plasmonic waves propagating along a chain of metallic nanoparticles in the presence of both a static magnetic field ${\bf  B}$ and a liquid crystalline host. The dispersion relations are obtained using the quasistatic approximation and a dipole-dipole  approximation to treat the interaction between surface plasmons on different nanoparticles.  For a plasmons propagating along a particle chain in a nematic liquid crystalline host and a field parallel to the chain, we find a small, but finite, Faraday rotation angle. In a cholesteric liquid crystal host and an applied magnetic field parallel to the chain, the dispersion relations for left- and right-moving waves are found to be different.  For some frequencies, the plasmonic wave may propagate only in one of the two directions.}

\begin{document}
\maketitle

\section{Introduction}

Ordered arrays of metal particles in dielectric hosts have many remarkable properties~\cite{meltzer,maier2,tang,park,pike2013,pike2013a}. For example, they support propagating modes which are linear superpositions of so-called "surface" or "particle" plasmons.   In dilute suspensions of metallic nanoparticles, these surface plasmons give rise to characteristic absorption peaks, in the near infrared or visible, which play an important role in their optical response~\cite{pelton,maier3,solymar,koederink,brong,maier03,park04,saj,ghenuche,alu06,halas,weber04,simovski05,abajo,jain,crozier}. For ordered arrays, if both the particle dimensions and the interparticle separation are much smaller than the wavelength of light, one can readily calculate the dispersion relations for both transverse ($T$) and longitudinal ($L$) waves propagating along the chain, using the quasistatic approximation, in which the curl of the electric field is neglected.

In a previous paper, we calculated these dispersion relations for metallic chains immersed in an anisotropic host, such as a nematic or cholesteric liquid crystal (NLC or CLC)~\cite{pike2013}.   Here, we consider the additional effects of a static magnetic field applied either parallel and perpendicular to the chain.   For a parallel magnetic field orientation we find that a linearly polarized $T$ wave undergoes a Faraday rotation as it propagates along the chain.  For a field of 2 Tesla, and reasonable parameters for the metal, this Faraday rotation  is around 1/2 a millidegree per ten interparticle spacings.  These results suggest that either a nematic liquid crystal (NLC) host or an applied magnetic field could be used as an additional ``control knob" to manipulate the properties of the propagating waves in some desired way. 

We also consider the same problem when the host is a cholesteric liquid crystal (CLC).   In this case, if the magnetic field is parallel to the chain and the director rotates about the chain axis with a finite pitch angle, we show that the frequencies of left- and right-propagating waves are, in general, not equal. This difference opens up the possibility that, for certain frequencies, a linearly polarized wave can propagate along the chain only in a single direction.  Thus, this geometry provides a possible realization of a one-way waveguide.  This realization is quite different from other proposals for one-wave waveguiding, such as electromagnetic waves propagating at the interface between two different materials~\cite{yu,dixit},  along chains of ellipsoidal particles in a spiral configuration~\cite{mazor,hadad}, and in arrays of gyromagnetic crystals~\cite{wang}.

The remainder of this article is organized as follows:  
First, we use the formalism of Ref.~\citenum{pike2013} to determine the dispersion relations for the $L$ and $T$ waves in the presence of an anisotropic host and a static  magnetic field.   Next, we give simple numerical examples and finally we provide a brief concluding discussion. 

\section{Formalism}
We consider a chain of identical metal nanoparticles, each a sphere of radius $a$, arranged in a one-dimensional periodic lattice along the $z$ axis, with lattice spacing $d$, so that the $n^{th}$ particle is assumed centered at $(0, 0, nd)$ ($-\infty < n < + \infty$).   The propagation of plasmonic waves along such a chain of nanoparticles has already been considered extensively for the case of isotropic metal particles embedded in a homogeneous, isotropic medium~\cite{brong}.  In the present work, we calculate, within the quasistatic approximation,  how the plasmon dispersion relations are modified when the particle chain is immersed in both an anisotropic dielectric, such as an NLC or CLC, and a static magnetic field. We thus generalize earlier work in which an anisotropic host is considered without the magnetic field~\cite{pike2013a,pike2013}.

In the absence of a magnetic field, the medium inside the metallic particles is assumed to have a scalar dielectric function.  If there is a magnetic field ${\bf B}$ parallel to the chain (which we take to lie  along the $z$ axis), the dielectric function of the particles becomes a tensor, $\hat{\boldsymbol \epsilon}_m$.   In the Drude approximation\cite{hui}, the diagonal components are $\epsilon(\omega)$,  while $\epsilon_{xy} = -\epsilon_{yx}= iA(\omega)$ and all other components vanish.  In this case, the components of the dielectric tensor take the form
\begin{equation}
\epsilon(\omega) = 1 - \frac{\omega_p^2}{\omega(\omega+ i/\tau )} \rightarrow  1 - \frac{\omega_p^2}{\omega^2},
\label{eq:epsw}
\end{equation}
and
\begin{equation}
A(\omega) = -\frac{\omega_p^2\tau}{\omega}\frac{\omega_c\tau}{(1-i\omega\tau)^2} \rightarrow \frac{\omega_p^2\omega_c}{\omega^3},
\label{eq:aw}
\end{equation}
where $\omega_p$ is the plasma frequency, $\tau$ is a relaxation time, and $\omega_c$ is the cyclotron frequency, and the second limit  is applicable when $\omega\tau \rightarrow \infty$.  We will use Gaussian units throughout.  
\begin{figure}[t]
\includegraphics[width=0.45\textwidth]{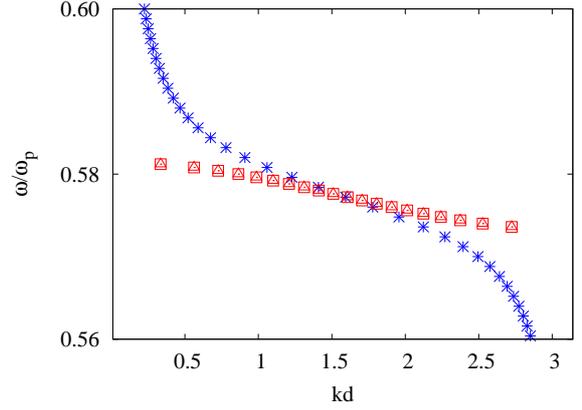} 
\caption{Blue symbols (x's and +'s): Dispersion relations for left and right circularly polarized $T$  plasmon waves propagating along a chain of metallic nanoparticles immersed in a NLC with both the director and a magnetic field parallel to a chain. The blue symbols correspond to metal particles are described by a Drude dielectric function with $\omega_p \tau = 100$ and $\omega_c/\omega_p = 3.5 \times  10^{-5}$.  The red symbols (open squares and triangles) represent the same dispersion relations as the blue symbols, but assuming no single-particle damping, corresponding to $\omega_p\tau = \infty$.   In both cases, the splitting between left and right circularly polarized waves is not visible on the scale of the figure. For $\omega_p = 1.0\times 10^{16}$ sec$^{-1}$, the chosen $\omega_c/\omega_p$ corresponds to $B \sim 2$ Tesla.}
\label{figure1}
\end{figure}

The dielectric function of the liquid crystal host, for either the NLC or CLC case, is taken to be that described in Ref.~\citenum{pike2013}.  The dispersion relations for the surface plasmon waves are determined within the formalism of Ref.~\citenum{pike2013}.   Specifically, we write down a set of self-consistent equations for the coupled dipole moments; these are given in Ref.~\citenum{pike2013} as Eq. (9), and repeated here for reference:
\begin{equation}
{\bf p}_n = -\frac{4\pi a^3}{3}\hat{\bf t}\sum_{n^\prime \neq n}\hat{\cal G}({\bf x}_n - {\bf x}_{n^\prime}){\bf p}_{n^\prime}.
\label{eq:selfconsist}
\end{equation}
Here 
\begin{equation}
\hat{\bf t} = \delta\hat{\boldsymbol{\epsilon}}\left(\hat{\bf 1}-\hat{\bf \Gamma}\delta\hat{\boldsymbol{\epsilon}}\right)^{-1}
\label{eq:tmatrix}
\end{equation}
is a ``t-matrix'' describing the scattering properties of the metallic sphere in the surrounding material, $\hat{\cal G}$ and $\hat{\bf \Gamma}$ are a $3\times 3$ Green's function and depolarization matrix given in Ref.~\citenum{pike2013}, $\hat{\bf 1}$ is the $3\times3$ identity matrix, and $\delta \hat{\boldsymbol{\epsilon}} = \hat{\boldsymbol{\epsilon}}_m - \hat{\boldsymbol{\epsilon}}_h$,  where $\hat{\epsilon}_h$ is the dielectric tensor of the liquid crystal host.   

\begin{figure}[t]
\includegraphics[width=0.45\textwidth]{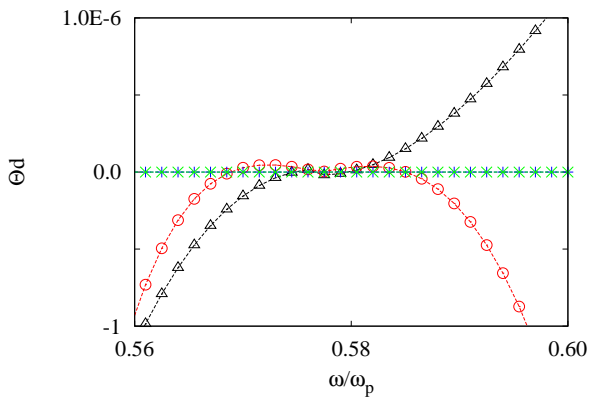} 
\caption{Real and imaginary parts of $\theta d$, the rotation angle per interparticle spacing (in radians), as a function of frequency, assuming $\omega_c/\omega_p = 3.5\times 10^{-5}$.  Nearly horizontal chains of blue ($+$) and green (x) symbols: Drude function for metal particles with no damping ($\tau \rightarrow \infty$).  Black triangles  and red  circles: finite damping ($\omega_p\tau = 100$).    In both cases, the magnetic field and the director of the NLC are assumed parallel to the chain axis, as in Fig.~\ref{figure1}. The dotted lines merely connect the points.}
\label{figure2}
\end{figure}

\subsection{Nematic Liquid Crystal}

We first consider a chain of metallic particles placed in an NLC host with ${\bf B}\| \hat{z}$ and parallel to the liquid crystal director $\hat{n}$.   Using the formalism of Ref.~\citenum{pike2013}, combined with Eq.~\eqref{eq:selfconsist}, we obtain two coupled sets of linear equations for the transverse (T) components of the polarization, $p_{nx}$ and $p_{ny}$.   The solutions are found to be left- and right-circularly polarized transverse waves with frequency $\omega$ and wave number $k_{\pm}$, where the frequencies and wave numbers are connected by the dispersion relations in the nearest-neighbor approximation 
\begin{equation}\label{eq:nlc_parallel_disp}
1= -\frac{2}{3} \frac{a^3}{d^3} \frac{\epsilon_\|}{\epsilon_\perp^2}\left(\frac{\epsilon(\omega)-1}{\epsilon(\omega)+2} \mp \frac{3 A(\omega)}{(\epsilon(\omega)+2)^2}\right)\cos(k_\pm d),
\end{equation}
where we use the notation of Ref.~\citenum{pike2013}. The longitudinal ($L$ or  $z$) mode is unaffected by the magnetic field.  Since the frequency-dependences of both $\epsilon(\omega)$ and $A(\omega)$ are assumed known, these equations represent implicit relations between $\omega$ and $k_\pm$ for these $T$  waves.

If ${\bf B} \|  \hat{x}$ while both $\hat{n}$ and the chain of particles are parallel to $\hat{z}$, then the dispersion relation for the coupled $y$ and $z$ waves are obtained  as solutions to the coupled equations
\begin{eqnarray}
p_{0y} & = &  \frac{-2a^3}{3d^3}\left[\frac{\epsilon_\|}{\epsilon_\perp^2}t_{yy}p_{0y} - \frac{2}{\epsilon_\perp}t_{yz}p_{0z}\right]\cos (kd), \nonumber \\
p_{0z} & = & \frac{-2a^3}{3d^3}\left[-\frac{\epsilon_\|}{\epsilon_\perp^2}t_{yz}p_{0y} -\frac{2}{\epsilon_\perp}t_{zz}p_{0z}\right]\cos (kd).
\label{eq:rotyz2}
\end{eqnarray}
These coupled $y$ and $z$ modes are uncoupled from the $x$ mode. 

If we solve this pair of equations for $p_{0y}$ and $p_{0z}$ for a given value of $k$, we obtain a nonzero solution only if the determinant of the matrix of coefficients vanishes. For a given real frequency $\omega$, there will, in general, be two solutions for $k(\omega)$ which decay in the $+z$ direction.   These correspond to two branches of propagating plasmon (or plasmon polariton) waves, with dispersion relations which we may write as $k_\pm(\omega)$.   The frequency dependence appears in $t_{yz}$, $t_{yy}$, and $t_{zz}$, which depend on $\omega$ [through $\epsilon(\omega)$ and $A(\omega)$].  However, unlike the case where the magnetic field is parallel to the  ${z}$ axis, the waves are elliptically rather than circularly polarized.

\subsection{Cholesteric Liquid Crystal}

We now consider immersing the chain of metallic nanoparticles in a CLC in the presence of a static magnetic field with ${\bf B}\| \hat{z}$ and the chain.  A CLC can be thought of as an NLC whose director axis lies perpendicular to a rotation axis (which we take to be $\hat{z}$), and which spirals about that axis with a pitch angle $\alpha$ per interparticle spacing.  For a CLC, if we include only interactions between nearest-neighbor dipoles, the coupled dipole equation [Eq.~\eqref{eq:selfconsist}] takes the form
\begin{equation}
{\tilde p}_n = -\frac{4\pi a^3}{3}[\hat{\bf R}^{-1}(z_1)\hat{\bf t}\hat{\cal G}\cdot{\tilde p}_{n+1} +\hat{\bf R}(z_1)\hat{\bf t}\hat{\cal G}\cdot{\tilde p}_{n-1}], 
\label{eq:selfcon_two}
\end{equation}
as is shown in Refs.~\citenum{pike2013} and~\citenum{pike2013a}.  Here $\tilde{p}_n = {\bf R}_n(z)p_n$ and ${\bf R}_n(z)$ is a $ 2\times2 $ rotation matrix for the director $\hat{n}(z)$. If the magnetic field lies along the $\hat{z}$ axis then the two $T$ branches are coupled.   One can write a  $2\times 2$ matrix equation for the coupled dipole equations in the rotated  ${x}$ and ${y}$ directions.  This equation is found to be
\begin{equation}
\tilde{\bf  p}_0 = -\frac{2 a^3}{3 d^3}\hat{\bf M}(k, \omega)\cdot \tilde{\bf p}_0,
\label{eq:disperchiral}
\end{equation}
where $\tilde{\bf p}_0$ is the rotated two-component column vector whose components are $\tilde{\bf p}_{x0}$ and $\tilde{\bf p}_{y0}$.   The components of the matrix $\hat{\bf M}(k,\omega)$ are found to be
\begin{eqnarray}
M_{xx} &=& \epsilon_1\lbrack t_{xx}\cos(kd)\cos(\alpha d) + it_{xy}\sin(kd)\sin(\alpha d)\rbrack \nonumber \\
M_{yy} &=& \epsilon_2\lbrack t_{yy}\cos(kd)\cos(\alpha d) + it_{xy}\sin(kd)\sin(\alpha d)\rbrack\nonumber \\
M_{xy} &=&\epsilon_2\lbrack t_{xy}\cos(kd)\cos(\alpha d) - it_{yy}\sin(kd)\sin(\alpha  d)\rbrack \nonumber \\
M_{yx} &=&\epsilon_1 \lbrack it_{xx}\sin(kd)\sin(\alpha d) - t_{xy}\cos(kd)\cos(\alpha d)\rbrack \nonumber. \\
\label{eq:magfield}
\end{eqnarray}
where $\epsilon_1 =  \frac{\epsilon_\perp^{1/2}}{\epsilon_\|^{3/2}}$ and $\epsilon_2= \frac{1}{\sqrt{\epsilon_\perp \epsilon_\|}}$.  One can now determine the dispersion relation for the two $T$ waves as non-trivial solutions to the secular equation formed from  Eqs.~\eqref{eq:disperchiral} and~\eqref{eq:magfield}.

 The most interesting result emerging from Eqs.~\eqref{eq:disperchiral} and~\eqref{eq:magfield} is that the presence of a magnetic field will lead to dispersion relations which are {\it non-reciprocal}, i.\ e., $\omega(k) \neq \omega(-k)$ in general.  The magnetic field appears only in the off-diagonal elements $t_{xy}$ and $t_{yx}$, which are linear in the field except for very large fields.  The terms involving $t_{xy}$ and $t_{yx}$ in Eq.~\eqref{eq:magfield} are multiplied by $\sin(kd)$ and thus change sign when $k$ changes sign. Thus, the secular equation determining $\omega(k)$ is not even in k, implying that the dispersion relations are non-reciprocal.  The non-reciprocal nature of the dispersion relations disappears at $B=0$ even though the off diagonal terms of ${\bf M}(k,\omega)$ are still nonzero, because $\sin(kd)$ appears only to second order.  Also, when the host dielectric is an NLC, the non-reciprocity vanishes because the rotation angle $\alpha =0$ and all terms proportional to $\sin(kd)$ vanish, even at finite $B$.

In a finite magnetic field, we define the difference in wave number between a right-moving or left-moving wave as 
\begin{equation} \label{eq:delta_k}
\Delta k_i(\omega) = \lvert \mathrm{Re}(k_{i,L})\rvert -\lvert\mathrm{Re}(k_{i,R})\rvert,
\end{equation}
where $i = 1,2$ for the two elliptical polarizations and $ L, R$ for either the left-moving or right-moving branch.  If, for example, $\Delta k(\omega) \neq 0$ then the left- and right-moving waves have different wave numbers for a given frequency and are non-reciprocal. 

\subsection{Faraday Rotation and Ellipticity}

By solving for $k(\omega)$ using either Eqs.~\eqref{eq:nlc_parallel_disp} or ~\eqref{eq:rotyz2} for an NLC, or~\eqref{eq:disperchiral} for a CLC,  one finds that the two modes polarized perpendicular to {\bf B} and  propagating along the nanoparticle chain have, in general,  different wave vectors.  For the NLC, we denote these $k_+(\omega)$ and $k_-(\omega)$, for the same frequency $\omega$, while for the CLC, we denote them $k_1(\omega)$ and $k_2(\omega)$. 

We first discuss the case of an NLC host and ${\bf B} \| \hat{z}$. Then the two solutions represent left- and right-circularly polarized waves propagating along the chain.  A linearly polarized mode therefore represents an equal-amplitude mixture of the two circularly polarized modes.  This mixture undergoes a {\it rotation} of the plane of polarization as it propagates down the chain and is analogous to the usual Faraday effect in a {\it bulk} dielectric.   The angle of rotation per unit chain length may be written
\begin{equation}
\label{eq:angle_single}
\theta(\omega) = \frac{1}{2} \left[k_+(\omega) - k_-(\omega)\right].
\end{equation}

\begin{figure}[t]
\includegraphics[width=0.45\textwidth]{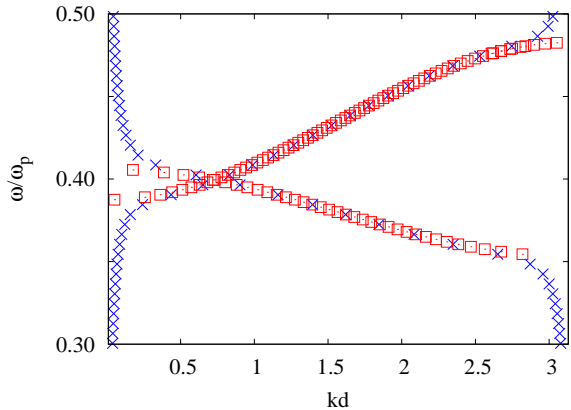} 
\caption{Dispersion relations for plasmon waves elliptically polarized in the $yz$ plane and propagating along a chain of metallic nanoparticles, assuming no damping (red open squares). The chain is assumed immersed in an NLC with director parallel to the chain ($\hat{z}$) with ${\bf B} \| \hat{x}$ and $\omega_c/\omega_p = 3.5 \times  10^{-5}$. Blue x's: same as red open squares, but assuming single-particle damping corresponding to $\omega_p\tau = 100$.      For $\omega_p = 1.0\times 10^{16}$ sec$^{-1}$, the chosen $\omega_c/\omega_p$ corresponds to about $ 2$ Tesla. }  
\label{figure3}
\end{figure}

In the absence of damping, $\theta$ is real.  If $\tau$ is finite, the electrons in each metal particle will experience damping within each particle, leading to an exponential decay of the plasmonic waves propagating along the chain. This damping is automatically  included in the above formalism, and can be seen most easily if only nearest neighbor coupling is included. The quantity
\begin{equation}
\theta(\omega) = \theta_1(\omega) + i\theta_2(\omega) 
\label{eq:thetatau}
\end{equation}
is then the {\it complex} angle of rotation per unit length of a linearly polarized wave propagating along the chain of metal particles.   By analogy with the interpretation of a complex $\theta$ in a homogeneous bulk material, $\mathrm{Re}\lbrack\theta(\omega)\rbrack$ represents the angle of rotation of a linearly polarized wave (per unit length of chain), and   $\mathrm{Im}\lbrack\theta(\omega)\rbrack$ is the corresponding Faraday ellipticity  i.\ e., the amount by which the initially linearly polarized wave becomes elliptically polarized as it propagates along the chain.  

\begin{figure*}[t]
\centering
\includegraphics[width=0.7\textwidth]{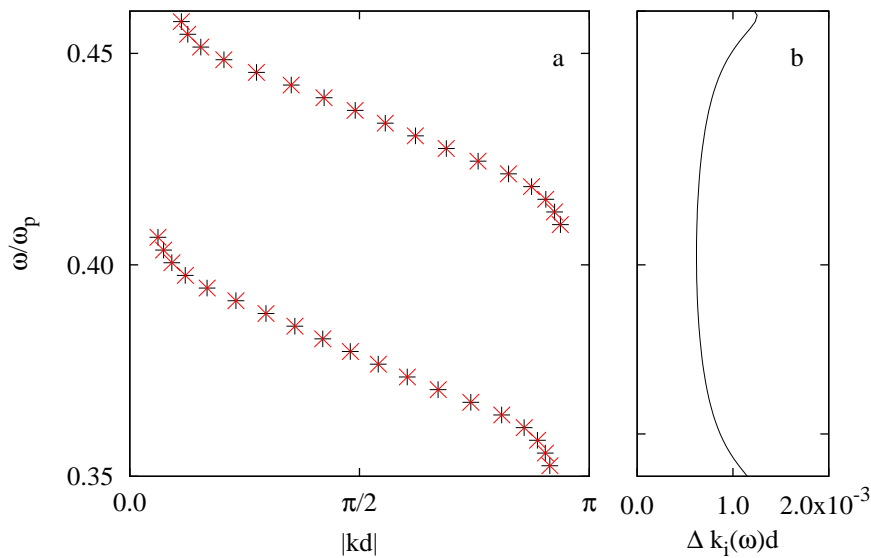} 
\vspace{-0.3in}
\caption{(a). Black $+$ symbols: the two dispersion relations for right-moving transverse plasmon waves propagating along a chain of metallic nanoparticles immersed in a CLC host with ${\bf B}\| \hat{z}$, plotted as a function of $\lvert kd \rvert$. Red x's: same quantities but for left-moving plasmon waves. We assume that $\omega_c/\omega_p = 3.5 \times  10^{-5}$, $\omega_p\tau = 100$, and $\alpha d = \pi/6$.  (b). The difference between the quantities $|k_id|$, i.\ e., $|\Delta k_id| = |Re(k_{i,L}d)| - |Re(k_{i,R}d)|$, for  left-propagating and right-propagating modes,  as given by Eq.~\eqref{eq:delta_k}, for the two elliptical polarizations. $L, R$ refer to the left-moving or right-moving waves, and $i$ (i = 1,2) labels the two branches for each direction.   Note that a non-zero value of $\Delta k_i(\omega)d$ implies that for a given frequency the left- and right-traveling waves have different wave vectors.  Numerically, we find that $\Delta k_i(\omega)d$ is independent of  $i$.}
\label{figure4}
\end{figure*}

In the case of a CLC host, neither of the two T modes is circularly polarized in general.  Thus, the propagation of a linearly polarized wave along the chain cannot be simply interpreted in terms of Faraday rotation.

\section{Numerical Illustrations}

We now numerically calculate the dispersion relations presented in the previous section using the liquid crystal known as E7.  This liquid crystal was described in experiments by M\"{u}ller~\cite{muller}, from which we take the dielectric constants $\epsilon_\|$ and $\epsilon_\perp$.  We first consider the case of an a NLC host with both the director and an
applied magnetic field parallel to the metallic chain axis $\hat{z}$. As noted earlier, the $L$ modes are unaffected by a magnetic field, but the $T$ modes are split into left- and right-circularly polarized waves.   To illustrate the predictions of our simple expressions, we take $a/d = 1/3$, and we assume a magnetic field such that the ratio $\omega_c/\omega_p = 3.5\times 10^{-5}$.   For a typical metallic plasma frequency of $\sim 10^{16}$ sec$^{-1}$, this ratio would correspond to a magnetic induction $B \sim 2 T$.   We consider both the undamped and damped cases; in the
latter, we choose $\omega_p\tau = 100$.  For propagating waves we choose solutions for which $\mathrm{Im}\lbrack k_\pm\rbrack > 0$ so that these waves decay to zero, as expected, when $z\rightarrow \infty$.

The calculated dispersion relations for the two circular polarizations of plasmonic wave are shown in Fig.~\ref{figure1} with and without single-particle damping.   The splitting between the two circularly polarized $T$ waves is not visible on the scale of the figure. In this, and all subsequent plots, we have calculated far more points than are shown in the Figure, so that effectively the entire range $0 < kd < \pi$ is included.

In Fig.~\ref{figure2}, we  plot the corresponding quantity $\theta d(k)$, the rotation angle
for a distance equal to one interparticle spacing,
 in Fig.~\ref{figure2}.    When there is no damping, we find that both the real and imaginary parts of  $\theta d$ are extremely small.  They become larger only when damping is included, as we do here by setting  $\omega_p\tau = 100$.  In this case,  neither $\mathrm{Re}\lbrack\theta(\omega)d\rbrack$ nor $\mathrm{Im}\lbrack\theta(\omega)d\rbrack$  exceed about $1\times10^{-6}$ radians, showing that a linear incident wave  acquires little ellipticity over such distances.   Since both theory and experiment suggest that the wave {\it intensity} typically has an exponential decay length of no more than around 20 inter-particle spacings in realistic metallic chains, the likely Faraday rotation of such a wave in practice will probably not exceed a millidegree or two, at most, even for a field as large as $2 T$.  Thus, while the rotation found here  may be measurable, it may not be large, at least for this simple chain geometry with one  particle per unit cell.    The present expressions also indicate that $\theta d$ is very nearly linear in $B$, so a larger rotation could be attained by increasing $B$. 

For a metallic chain in a NLC where $B \perp \hat{ z}$, we find, using the same parameters and requirements as the previous case, that the two non-degenerate waves (one an $L$ and the other a $T$ wave) become mixed  when $B\neq 0$.  The dispersion relations, again with and without damping, are plotted in Fig.~\ref{figure3}.  When compared to previous work in Ref.~\citenum{pike2013}, the dispersion relations in Fig.~\ref{figure3} are modified because of the finite damping and presence of the magnetic field.  However, the change produced by the magnetic field in these dispersion relations is not visible in the figures.

Finally, we discuss the case of metallic chain parallel to the $z$ axis, subjected to a magnetic field along the $z$ axis, and immersed in a CLC whose twist axis is also parallel to $\hat{z}$.  Using the same host dielectric constants given above and a twist angle of $\alpha d = \pi/6$,  we show in Fig.~\ref{figure4}(a) the resulting dispersion relations, i.\ e.,  $\omega/\omega_p$ plotted against $\lvert kd \rvert$, for the two transverse branches.  In particular, we show both transverse branches for a right-moving wave (black symbols) and left-moving wave (red symbols) giving a total of $4$ plots shown in Fig.~\ref{figure4}(a).   The separation between the two $T$ branches is on the order of $0.05 \  \omega/\omega_p$ for all$k$.  In Fig.~\ref{figure4}(b), we plot the corresponding difference in wavenumber between the left- and right-moving waves as $\Delta k_i(\omega)d$.  For the parameters used, this difference turns out to be the same, to within numerical accuracy, for both the transverse branches, and hence we show only a single plot.  Since $\Delta k_i(\omega)d$ is nonzero in a wide frequency range, the wave propagation is indeed non-reciprocal in this range.  One-way wave propagation may occur in part of this  range.  Such propagation would occur when, at particular frequencies, waves can propagate only in one of the two directions.   We have not, however, checked this possibility numerically.

\section{Discussion}

The present numerical calculations omit several potentially important factors which could alter the numerical results. Among these are the effects of particles beyond the nearest neighbors on the dispersion relations~\cite{brong}, the (possibly large) influence of the particles in disrupting the director orientation of the liquid crystalline host~\cite{lubensky,poulin,stark,kamien,allender}, and the effects of radiative damping~\cite{weber04} on the dispersion relations.  Nonetheless, we believe that  our calculations correctly describe, at least qualitatively, how the surface plasmon dispersion relations are affected by the combined influence of a liquid crystalline host and an applied magnetic field.

It should be noted that the magnetic field effects described in this paper are numerically very small, for the parameters investigated.  The smallness is caused mainly by the small value of the ratio $\omega_c/\omega_p$, taken here as $3.5 \times 10^{-5}$.   To increase this ratio, one could either increase $\omega_c$ (by raising the magnetic field strength), or decrease $\omega_p$ (by reducing the free carrier density in the metal particle).  We speculate this latter effect could be achieved by studying n-type semiconductor nanoparticles, if these could be prepared in a free carrier regime without having to consider quantized energy levels.   The same calculations should also be carried out for more realistic metallic dielectric functions than the Drude model.

In summary, we have calculated the dispersion relations for plasmonic waves propagating along a chain of metallic nanoparticles immersed in a liquid crystal and subjected to an applied magnetic field.   For a magnetic field parallel to the chain and director axis of the NLC, a linearly polarized wave is Faraday-rotated by an amount proportional to the magnetic field strength.   For a CLC host and a magnetic field parallel to the chain, the transverse wave solutions become non-reciprocal (left and right-traveling waves having different dispersion relations) and there may be frequency ranges in which waves can propagate only in one direction.   Thus, plasmonic wave propagation can be tuned, either by a liquid crystalline host or a magnetic field, or both. In the future, it may be possible to detect some of these effects in experiments, and to use some of the predicted properties for applications, e. g., in optical circuit design.

\section{Acknowledgments}

This work was supported by the Center for Emerging Materials at The Ohio State University, an NSF MRSEC (Grant No.\ DMR-1420451).

\end{document}